\documentclass[conference]{IEEEtran}

\usepackage{cite}
\usepackage{underscore}
\usepackage[english]{babel}
\usepackage{blindtext}
\usepackage{lipsum}
\usepackage{algorithmic}
\usepackage{graphicx}
\usepackage{xcolor}

\usepackage[tight,footnotesize]{subfigure}
\usepackage{caption}
\usepackage{xcolor}
\ifCLASSINFOpdf
\else
\fi

\usepackage{etoolbox}
 \makeatletter
 \patchcmd{\@maketitle}
   {\addvspace{0.5\baselineskip}\egroup}
   {\addvspace{-0.5\baselineskip}\egroup}
   {}
   {}
 \makeatother

\begin{document}
%
\title{Cache peering in multi-tenant 5G networks}

\author{
\IEEEauthorblockN{Konstantinos V. Katsaros, Vasilis Glykantzis, George Petropoulos}%
\IEEEauthorblockA{Intracom SA Telecom Solutions, \\
Peania, 19002, Greece\\
Email:\{konkat, vasgl, geopet\}@intracom-telecom.com}%
}


%


\maketitle

\begin{abstract}
Building on the adoption of the Network Functions Virtualization (NFV) and Software Defined Networking (SDN) paradigms, 5G networks promise distinctive features including the capability to support \textit{multi-tenancy}. Virtual network operators (VNOs) are expected to co-exist over the shared infrastructure, realizing their network functionality on top of virtualized resources. In this context, we observe the emerging opportunity for establishing synergies between co-located tenants of the infrastructure, in the form of cache peering relationships between co-located VNOs. Upon a cache miss, co-located caches benefit from content cached at their peers, taking advantage of the shared nature of the infrastructure in reducing latencies and traffic overheads. Our approach allows VNOs to autonomously manage their peering links without the involvement of the infrastructure operator.
\end{abstract}


%
\IEEEpeerreviewmaketitle

\section{Introduction}\label{intro}

5G networks are expected to adopt the NFV paradigm, opening the way to a series of benefits. By realizing key network functions (e.g., caches, firewalls, IDS, DPI, etc.) on top of virtualized compute, storage and network resources, NFV promises a series of benefits such as the reduction of associated CAPEX/OPEX, due to the shared nature of the equipment, as well as the increased flexibility in network management and programmability. 
In the context of 5G, these still shaping capabilities, facilitate a series of key features and advances, including the assembly and management of isolated sets of virtual resources (often termed as \textit{network slices}) tailored for the operational needs of third parties i.e., the tenants of the shared infrastructure. In turn, this capability supports the emergence of Virtual Network Operators (VNOs), enabling multi-tenancy scenarios, where multiple VNOs are realized on top of the same, shared physical infrastructure. 

\begin{figure}[ht!]
\centering
\subfigure[Cache miss: no cache peering]{
\includegraphics[width=3.5in]{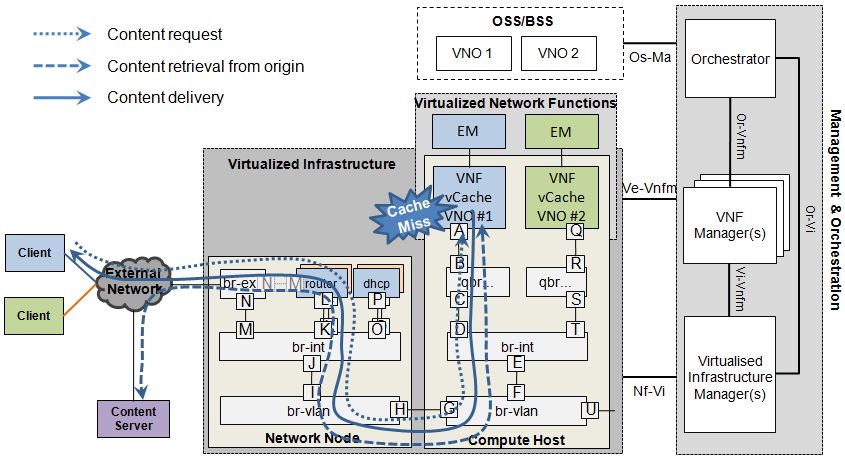}
\label{cache_miss_no_peering}
}
\subfigure[Cache miss: peering with co-located cache]{
\includegraphics[width=3.5in]{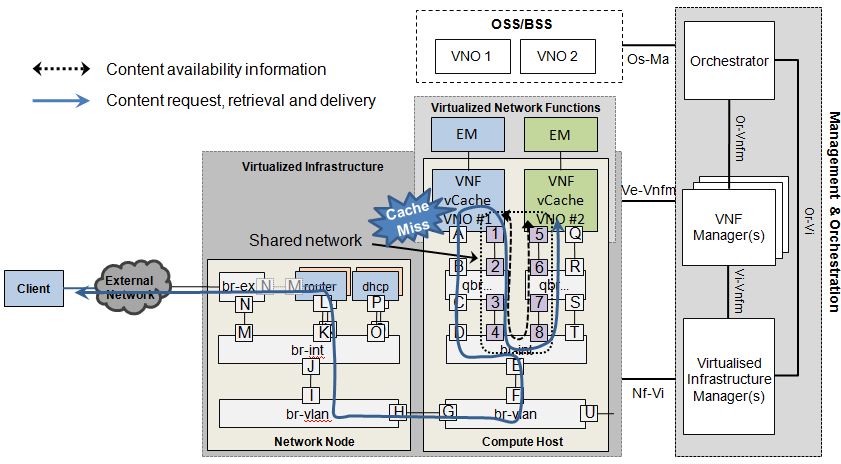}
\label{cache_miss_peering}
}
\caption[]{Caching in NFV: cache misses and the effect of cache peering}
\label{caching_example}
\end{figure}

Typically, communication between tenants is expected to take place by traversing the corresponding mobile network gateways, since VNOs will be realized as entirely distinct networks. However, the shared nature of the underlying infrastructure presents the opportunity for shortcuts in this communication. As network functions of different VNOs may reside on the same IT infrastructure i.e., micro-data center ($\mu$-DC), inter-VNO traffic can be exchanged locally, thus promising significant benefits in terms of end-to-end latency and overall traffic overheads
\footnote{This operation raises challenges related to the GTP tunnels used to deliver traffic to the mobile network gateway. Potential workarounds have been proposed in literature \cite{Rodrigues2016}. This issue is consider out of this paper's focus area, which is centered around the particular issue of cache peering.}. Building on this observation, we focus on the particular case of caching, proposing the establishment of cache peering relationships between \textit{co-located} VNOs. In the envisioned environment, VNOs instantiate transparent virtual caches (vCaches) to reduce traffic overheads and improve performance for their users. Co-location takes the form of vCaches instantiated within the same $\mu$-DC, either on the same or different compute hosts (i.e., physical servers). Cache peering then takes advantage of the proximity of vCaches i.e., upon a local cache miss, peering vCaches mutually benefit from the content cached at co-located VNOs. As illustrated in Figure\ref{cache_miss_no_peering}, in the absence of cache peering, a cache miss results in the traversal of the virtualized infrastructure (VI) 4 times, before the content can be delivered to the end user i.e., a content request exits the VI to reach the content origin, bringing the content to the vCache before exiting the VI again to reach the end user. In contrast, cache peering allows content to be locally fetched from a co-located vCache, promising substantial traffic and latency savings. What is more, 
the inherently low latency/high bandwidth $\mu$DC environments are expected to facilitate communication between peering vCaches, including both the exchange of content availability information and the delivery of content. 

The envisioned setup contributes to the emergence of new business models, for the NFV-enabled cooperation between VNOs, in an analogy to inter-domain peering agreements between autonomous systems in the Internet~\cite{Labovitz2010}. However, the considered environment presents increased complexity. Cache peering involves the implicit sharing of compute and storage resources, in addition to network resources. Moreover, VNOs are expected to pursue their autonomy in managing their (business) peering relationships, consequently calling for   technical solutions that do not require the intervention/involvement of the (third party) infrastructure operator. 

In this paper, we report on our ongoing work towards the realization of cache peering relationships in the aforedescribed context. We present technical details for the configuration of the proposed solution for a typical OpenStack-supported $\mu$-DC environment, identifying key challenges related to virtualization overheads, network slice traffic isolation, as well as the tight control of resource consumption. 



\section{Background}\label{background}

\subsection{NFV and multi-tenancy}

The NFV concept has been under intense investigation by the research community during the last years, largely building on a series of architecture specifications produced by the ETSI standardization body\cite{ETSINFV}, as well as a series of open source tools e.g., Open Source Mano (OSM)\footnote{https://osm.etsi.org/}. Figure \ref{caching_example} provides an illustration of this architecture. The management and orchestration (MANO) plane provides infrastructure operators with the capability to manage their virtualized infrastructure (VI), allocating resources for the realization of virtual network functions (VNFs), in the form of virtual machines (VMs), and interconnecting them with the rest of the (virtual) network infrastructure, into end-to-end network services. In the context of 5G multi-tenancy support, these services span the entire network and IT infrastructure of the infrastructure operator, comprising the network slices allocated to each instantiated VNO.  


Taking a closer look at the VI, and in order to understand the technical implications of MANO operations in the targeted environment, we adopt a typical\footnote{OpenStack configuration is rather complex, offering a series of features and options. We provide a simplified description so as to improve clarity of presentation, within length limitations. For an elaborate description the reader is first referred to http://docs.openstack.org/liberty/networking-guide/index.html} OpenStack configuration, also shown in Figure \ref{caching_example}. The IT and network resources of the VI are organised in Compute and Network hosts/nodes. Compute nodes (or hosts) host VNFs in the form of VMs. Network nodes are responsible for handling ingress/egress traffic, isolating and optimising traffic flow across tenants. This is accomplished by a series of Layer 2/3 forwarding nodes (e.g., Open vSwitch instances (OVS)\footnote{openvswitch.org/}). \texttt{qbr} switches are introduced to handle security issues, while \texttt{br-ex} is responsible for handling traffic from/to the external network. The \texttt{br-int} switch has a central role in switching packets on a node level. The \texttt{br-vlan} switch is responsible for handling traffic labeled with 802.1Q VLAN tags. A separate VLAN ID can be used for each VNO to isolate traffic at Layer 2. 

\subsection{Caching}

Content caching has been intensively investigated in the past, with a particular focus on Web (HTTP) caching (e.g., \cite{Wang1999}). Caches are a typical part of networks and/or content delivery networks (CDNs) in the form of middleboxes/dedicated servers. Traffic is typically intercepted by caches via means of user request redirection, either in the form of DNS redirections, in the case of CDNs, or via the end user proxy cache configuration.
 Both cases are associated to penalties related to the DNS signaling latencies and complexity and/or the cumbersome configuration of user devices. 
 In an alternative approach, transparent caching is building on the network configuration, to forward traffic to a cache instance. 

Taking advantage of distributed cache resources, cooperative caching schemes have been developed, enabling the exchange of cached content by remote caches. This exchange is realized with the establishment of relationships between the caches including \textit{cache peering} relationships (alternatively, peering caches are known as  \textit{siblings}). When a content request results in a cache miss i.e., the content is not locally available, a cache can direct the request to a sibling cache. The receiving sibling cache responds either with the content or with a failure message. Sibling caches indicate the availability of content to each other, either proactively e.g., exchanging compact representations of their cached contents\footnote{E.g., Cache Digests, http://wiki.squid-cache.org/SquidFaq/CacheDigests} or reactively (i.e., upon a request) by directly querying each other \cite{rfc2186}. 


\section{Cache Peering for 5G tenants}\label{architecture}

\subsection{Baseline tenant configuration}

Enabling cache peering in an NFV-enabled multi-tenant environment, first goes through the end-to-end network service orchestration, including the realization of the caching service at each tenant separately. This first involves the configuration of the network so as to ensure the isolation of tenant traffic. To this end, we consider the  establishment of a VLAN segment per VNO. Edge network devices (on the access network) are configured by the MANO plane to tag the incoming traffic with a VLAN ID allocated to the VNO. All involved forwarding elements in the infrastructure e.g., backhaul links, are also correspondingly configured to associate VNO traffic (and allocated resources e.g., bandwidth) with the specific VLAN ID. 
The MANO plane subsequently orchestrates the instantiation of the vCaches, allocating the requested IT resources e.g., CPU, RAM, storage, etc. In this stage, a separate tenant is created within the VI i.e., OpenStack in our case. This results in the secure isolation of allocated resources across tenants of the VI, including network isolation i.e., a separate virtual network is created for each tenant within the VI in the form of a VLAN. At this point, the Orchestrator component is responsible for coordinating the overall network configuration, so that the entire forwarding fabric (within and across $\mu$DC borders), appears as a single VLAN domain. 


\subsection{Traffic interception}

The next stage involves the configuration of the network for the interception of traffic by the vCaches. As the cache VMs are configured to participate in the tenant's VLAN domain, all traffic can reach them using well known techniques (see Section \ref{background}). However, the overheads associated with a cache miss in the context of NFV caching, as illustrated in Section \ref{intro}, call for a more agile approach, where vCaches are reached mainly/only by flows that are likely to result in a cache hit. The emergence of SDN has introduced the missing agility and flexibility in dynamically identifying the traffic to be intercepted by caches, and appropriately steering traffic through flow rules, subject to content availability and load conditions \cite{Rodrigues2016, Kimmerlin2014, Georgopoulos2014}. However, realizing these solutions in an NFV-enabled, multi-tenancy environment requires careful consideration of the shared nature of the infrastructure. Intelligent traffic interception mechanisms, currently build on content availability lookups upon each content request. As such, they are prone to significant resource consumption of switching fabric, such as \texttt{br-int} at the network node, raising concerns regarding the overall forwarding performance, even for VNOs with no vCaches in operation. Alternatively realizing these solutions on a VNF level, i.e, allocating VM resources for the content availability lookups,  would confine the impact within VNO resources. However, a close look at Figure \ref{caching_example} illustrates the drawback of this approach: traffic steering decisions are taken once traffic flows have already traversed the virtualized infrastructure towards the VNF; flows that will eventually bypass the vCache, pay the corresponding traversal delay penalties. 

Considering this tradeoff, our on-going work focuses on the establishment of traffic interception flow rules on the shared \texttt{br-int}, however avoiding the aforementioned lookup operations and the associated overheads. Instead, the definition of interception rules relies on the pro-active processing of vCache access logs/cache index, with the purpose of identifying target IP addresses prone to cache misses e.g., popular web sites serving personalized content. Our future work plans include a detailed assessment of this design, focusing on the accuracy of the traffic interception rules, their memory footprint and impact on lookup and latency savings.


\subsection{Peering configuration}

Based on the aforementioned configurations, each VNO is in the position to provide caching support within its network slice. The established network configuration does not allow traffic to cross VNO borders e.g., an HTTP request of a user in VNO A can never reach any VNO B vCache, and both VNO's vCaches cannot communicate with each other, even if they are instantiated within the same Compute Host. The establishment of a cache peering relationship, then calls for the careful configuration of the network environment, adhering to the following set of requirements. The main objective is to allow peering vCaches to communicate so as to exchange content availability information, cache requests and content (Req.1). The provided solution however should not allow any other form of traffic to traverse the inter-VNO communication link (Req.2). The reason is that a peering agreement requires a well defined interface, over which the aforementioned control and data plane peering traffic is solely exchanged, avoiding misuse of the established communication link/network i.e., VNOs should not be allowed to directly offload user traffic to peering vCaches, so as to reduce local resource consumption. In the same vein, the authentication/authorization of the involved vCaches is required to mitigate any hijacking of the peering communication link/network from malicious third parties i.e., malicious tenant exploiting VNO network configuration vulnerabilities to make unauthorised use of the peering vCache resources (Req.3). At the same time, the communication between the peering vCaches requires a low latency and high bandwidth communication link/network so as to avoid delay penalties in the discovery and delivery of the requested content from peering vCaches, thus preserving the key benefits of and motivation for peering (Req.4).   

Towards these ends, the proposed solution includes a mixture of network and application level solutions. On the network side, our approach foresees the creation of a shared network between the involved vCaches. This network is shared exclusively by the vCache instances i.e., the VNF VMs of the involved tenants. The  Role-Based Access Control (RBAC)\footnote{http://docs.openstack.org/liberty/networking-guide/adv-config-network-rbac.html} feature introduced in Liberty version of OpenStack, enables tenants to grant access to network resources for specific other tenants. Building on this feature, a VNO first creates a network instance, further also configuring its own vCache(s) by adding to them an interface to the new network. Subsequently, the VNO is able to grant access to the peering VNO by its tenant ID. The latter VNO is able then to proceed with attaching its vCache(s) to the shared network. When the vCaches are collocated on the same Compute Host, the resulting configuration allows the peering traffic to reach the corresponding VMs by only traversing the \texttt{br-int} switch (see  Figure \ref{cache_miss_peering}). When vCaches are located at different Compute Hosts of the same $\mu$DC, traffic reaches the host of the peering vCache through a direct link typically available between \texttt{br-vlan} switches of co-located Compute Hosts (not shown due to length limitations). 

The described configuration so far satisfies Req. 1 and 4. To further satisfy Req. 2 and 3, application level configurations come into play. Namely, vCaches build on the Squid cache implementation\footnote{http://www.squid-cache.org/}, a mature and widely adopted solution. The configuration of Squid first includes the establishment of the peering link, through the \texttt{cache_peer} directive\footnote{http://www.squid-cache.org/Doc/config/cache_peer/}, which allows the specification of the peering vCache IP/hostname and listening ports. Satisfying Requirement 2, goes through the \texttt{iptables} configuration of a vCache, dropping all not legitimate input traffic e.g., traffic destined to a non-peering port or originated by an IP address other than the vCache IP address\footnote{A malicious peering VNO can still though offload all its HTTP traffic.}. Satisfying Requirement 3, goes through the appropriate \texttt{login} configuration options of the \texttt{cache_peer} Squid directive. The configuration of the \texttt{icp_access} access control list, enables access control of cache peering control traffic.

It is noted that the proposed configuration does not rely on the direct involvement of the infrastructure operator i.e., both the establishment of the shared network and the application layer configuration are carried out by the tenants themselves
. This is considered as an important aspect of the solution, simplifying and facilitating the management of peering links by the VNOs, without directly exposing business relationships to third parties.

As the establishment of a vCache peering link realizes a business agreement between the VNOs, monitoring and controlling the amount of \textit{exchanged} resources becomes particularly important, as it allows VNOs to ensure the symmetry of the peering link in terms of resource consumption. This means that VNOs require firm control over the compute, storage and network resources devoted to the peering agreement (Req.5). In the context of caching, this translates to the compute load for the lookup of content for peering requests, the I/O storage load for the retrieval of cached content and the load for the transmission of the content to the peering vCache. Our preliminary approach on this issue, builds on the Delay Pools feature of the Squid cache implementation, which provides the means to limit the bandwidth of certain requests based on any list of criteria\footnote{http://wiki.squid-cache.org/Features/DelayPools}. In particular, we define a class 2 delay pool for the shared network between the vCaches.



\section{Related Work}

Caching has been identified as one of the key applications areas for the NFV paradigm, right from the beginning of the NFV concept\cite{NFV2012}. Since then, several commercial NFV-enabled solutions have appeared, including caching as a key building block of a broader-CDN oriented solution e.g., \cite{Qwilt2014, JuniperAkamai2014}. However, limited information has been revealed about the service configuration, as these solutions are proprietary, while, to the best of our knowledge, no commercial solution focusing on multi-tenancy. 
The currently on-going 5G PPP Phase 1 EU H2020 research projects put substantial effort in integrating and enhancing the NFV paradigm within the 5G landscape \cite{5GPPP}. However, again, to the best of our knowledge, no research efforts are being devoted to the particular case of cache peering in multi-tenancy environments. 
It is also worth noting that, in the recent past, a series of efforts have been devoted in enabling the extension of (caching) service footprint through peering, in the context of Content Delivery Network interconnection (CDNi) \cite{rfc6770}; however, these efforts aimed at the design of interfaces between (application level) CDNs, rather than building on the emerging NFV capabilities and the overall integration of IT resources within the 5G network infrastructure. In all, we consider the proposed cache peering approach as an important step in progressing beyond the mere NFV-based instantiation of virtualized caches, focusing on a better understanding of the particular technical (and business) opportunities brought in the field by virtualization and network programmability, namely, for multi-tenancy.

\section{Conclusions and Future Work}

Building on NFV and SDN technical advances, and the emerging capability to support multi-tenancy in 5G networks, in this paper we proposed the realization of cache peering relationships between VNOs. The purpose is to take advantage of the potential co-location of virtual caches within $\mu$DCs, thus promising reduced latency and traffic overheads. We presented our on-going work, identifying key challenges and potential solutions in the specific context of NFV. The proposed technical path towards the realization of the envisioned setup enables the autonomous management of peering links from VNOs.






%
\bibliographystyle{IEEEtran}
\bibliography{refs}

\end{document}